\documentclass[pra,twocolumn]{revtex4-1}
\usepackage{hyperref} 

\usepackage{graphicx} 
\usepackage{subfig}
\usepackage{color}
\usepackage{filecontents}
\usepackage{soul}
\usepackage{mathrsfs}
\usepackage{tikz,bm} 
\usepackage{circuitikz}
\usepackage{tikz}
\usepackage{physics}
\usepackage{mathtools}
\usepackage[T1]{fontenc}
\usepackage[outline]{contour}
\usepackage{relsize}

\usetikzlibrary{shapes.geometric}
\usetikzlibrary{decorations.pathmorphing}
\usetikzlibrary{shapes.arrows, fadings}

\definecolor{mycolor}{rgb}{1,0.2,0.3}
\definecolor{brightgreen}{rgb}{1.0, 1.0, 1.0}
\definecolor{britishracinggreen}{rgb}{0.0, 0.26, 0.15}
\definecolor{cadmiumgreen}{rgb}{0.0, 0.42, 0.24}
\definecolor{ceruleanblue}{rgb}{0.16, 0.32, 0.75}
\definecolor{darkelectricblue}{rgb}{0.33, 0.41, 0.47}
\definecolor{darkpowderblue}{rgb}{0.0, 0.2, 0.6}
\definecolor{dt}{rgb}{1.0, 0.66, 0.07} 
\definecolor{emerald}{rgb}{0.31, 0.78, 0.47}
\definecolor{palatinatepurple}{rgb}{0.41, 0.16, 0.38}
\definecolor{pastelviolet}{rgb}{0.8, 0.6, 0.79}
\definecolor{br}{rgb}{0.5, 0.05, 0.01}
\definecolor{chosen_color}{RGB}{3, 207, 252}

\pgfdeclareverticalshading{pgf@lib@fade@north}{100bp}
{color(0bp)=(pgftransparent!0);
 color(30bp)=(pgftransparent!0);
 color(50bp)=(pgftransparent!80); 
 color(65bp)=(pgftransparent!100);
 color(100bp)=(pgftransparent!100)}%
\pgfdeclarefading{exponentialtop}{%
\pgfuseshading{pgf@lib@fade@north}%
}

\pgfdeclareverticalshading{pgf@lib@fade@north}{100bp}
{color(0bp)=(pgftransparent!100);
 color(35bp)=(pgftransparent!100);
 color(50bp)=(pgftransparent!80); 
 color(70bp)=(pgftransparent!00);
 color(100bp)=(pgftransparent!00)}%
\pgfdeclarefading{exponentialbottom}{%
  \pgfuseshading{pgf@lib@fade@north}%
}

\newcommand{\be}{\begin{equation}}
\newcommand{\ee}{\end{equation}}
\newcommand{\bea}{\begin{eqnarray}}
\newcommand{\eea}{\end{eqnarray}}
\newcommand*{\myeqref}[2][Eq.~]{%
\hyperref[{#2}]{#1(\ref*{#2})}%
}
\def\equationautorefname#1#2\null{%
Eq.#1(#2\null)%
}

\usepackage{dcolumn}
\usepackage{bm}
\usepackage{amssymb,amsmath}
\usepackage{color}
\usepackage{float}
\usepackage{tikz}
\usetikzlibrary{arrows}

\definecolor{DarkGreen}{rgb}{0,0.6,0.2}

\begin{document}
\title{Engineering Optomechanically Induced Transparency by coupling \\a qubit to a spinning resonator}
\author{Jessica Burns$^{1,3,\ast}$, Owen Root$^{2,3,}$}
\thanks{These authors have contributed equally to this work.}
\author{Hui Jing$^{3,4}$}
\author{Imran M. Mirza$^{5}$}
\email{mirzaim@MiamiOH.edu}
\affiliation{$^1$Physics Program, University of Cincinnati, OH 45221, USA\\
$^2$Physics Program, Nebraska Wesleyan University, Lincoln, NE 68504, USA\\
$^3$Key Laboratory of Low-Dimensional Quantum Structures and Quantum Control of Ministry of Education,\\
Department of Physics and Synergetic Innovation Center for Quantum Effects and Applications,\\
Hunan Normal University, Changsha 410081, China\\
$^4${Synergetic Innovation Academy for Quantum Science and Technology}, {Zhengzhou University of Light Industry, Zhengzhou 450002, China}\\
$^5$Department of Physics, Miami University, Oxford, OH 45056, USA}
\date{\today}

\begin{abstract}
We theoretically study the spectral properties of a pump-probe driven hybrid spinning optomechanical ring resonator optically coupled with a two-level quantum emitter (QE or qubit). Recently we have shown [Optics Express, 27, 18, 25515--25530 (2019)] that in the absence of the emitter the coupled cavity version of this setup is not only capable of nonreciprocal light propagation but can also exhibit slow \& fast light propagation. In this work, we investigate in what ways the presence of a single QE coupled with the optical whispering gallery modes of the spinning optomechanical resonator can alter the probe light nonreciprocity. Under the weak-excitation assumption and mean-field approximation, we find that the interplay between the rotational/spinning Sagnac-effect and the qubit coupling can lead to the enhancement both in the optomechanically induced transparency (OMIT) peak value and in the width of the transparency window due to the opening of qubit-assisted back reflection channel. However, compared to the no-qubit case, we notice that such an enhancement comes at the cost of degrading the group delay in probe light transmission by a factor of 1/2 for clockwise rotary directions. The target applications of these results can be in the areas of quantum circuitry and in non-reciprocal quantum communication protocols where QEs are a key component.
\end{abstract}

\maketitle


\section{Introduction}
Bulk Faraday rotators, either based on magneto-optical crystals (for instance Yttrium Aluminum Garnet-YAG) \cite{stadler2013integrated} or alkali vapor cells (such as Rb) \cite{siddons2010optical} present a key example of nonreciprocal optical devices. Commercially available nonreciprocal optical elements, for example, the ones with $45^0$ shifting in the polarization plane require Verdet constants $V(\lambda)$ of almost $80$ $rad/Tm$ for a centimeter-long crystal when a magnetic field of $1T$ is applied parallel to the propagation direction of electromagnetic radiation. However, when such an element is brought into smaller scales for quantum photonics applications, due to magnetic field strength limitations, it turns out that even for a  $100\mu m$ long crystal at $1T$ a Verdet constant of $8000$ $rad/Tm$ is required to achieve $45^0$ shifting. Unfortunately, not all magneto-optical crystals or Alkali vapors are capable of demonstrating such a high value of $V(\lambda)$ for a wide range of wavelength/$\lambda$ values. The matter is further worsened by the fact that even if such a high value of $V(\lambda)$ is attained for certain $\lambda$ values it is achieved at the price of higher losses. These considerations pose severe challenges to incorporating traditional nonreciprocal elements in integrated quantum photonics.

\begin{figure*}
\centering
  \begin{tabular}{@{}cccc@{}}
   \includegraphics[width=4.5in, height=3.25in]{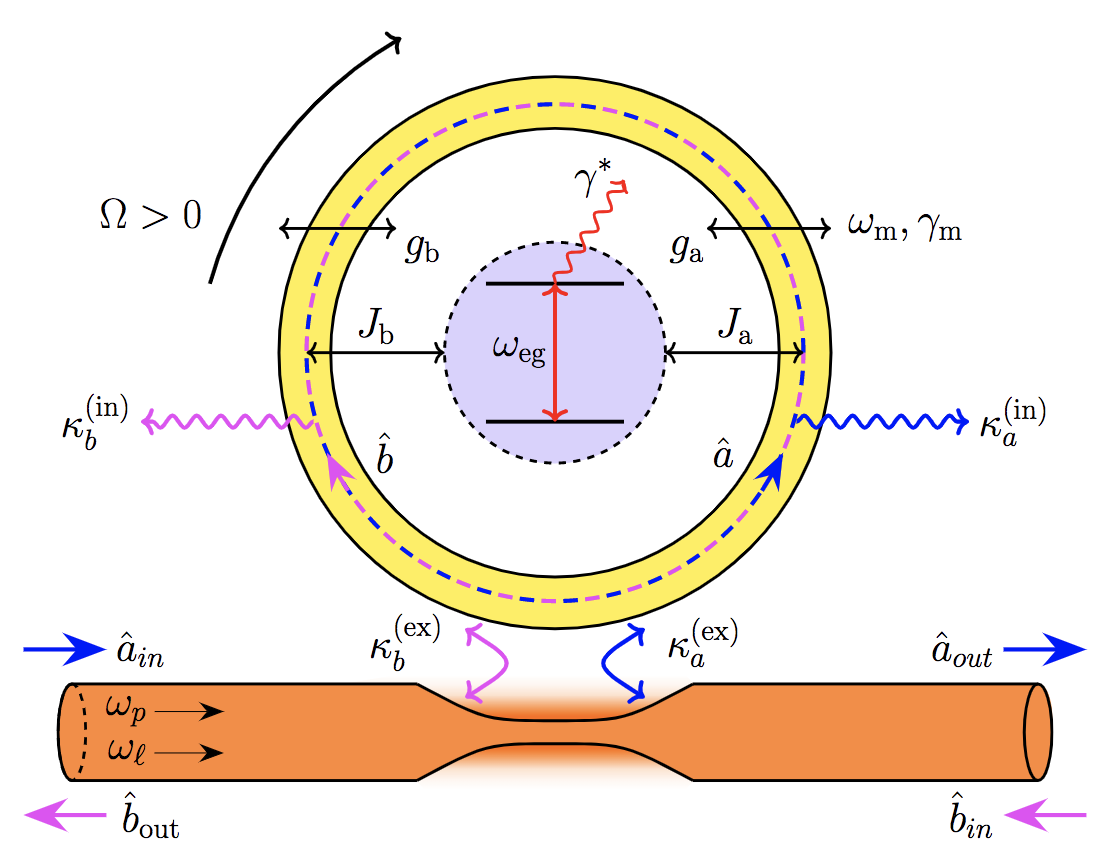}
 \end{tabular}
\captionsetup{
  format=plain,
  margin=1em,
  justification=raggedright,
  singlelinecheck=false
}
 \caption{(Color online) Model for the hybrid spinning atom-optomechanical microresonator architecture considered in this paper. In the right channel of the tapered fiber, $\hat{a}_{in}$ and $\hat{a}_{out}$ respectively represent the probe field input and output operators. Whereas, in the left channel of the fiber (which is assumed to be driven by a vacuum field) the input and output operators are given by $\hat{b}_{in}$ and $\hat{b}_{out}$, respectively. For further details about the system see Sec.~II(A).}\label{Fig1}
\end{figure*}
To address these issues, in recent years, hybrid quantum systems \cite{kurizki2015quantum, xiang2013hybrid, kotler2017hybrid, wallquist2009hybrid, mirza2016strong, mirza2015real} have emerged as a potential solution. For instance, in 2013 Peng et al. studied non-reciprocity in light transmission by breaking the $\mathscr{P}\mathscr{T}$-symmetry in on-chip coupled microtoroid resonators \cite{peng2014parity}. Extending this work to the hybrid domain, Zheng et al. considered two coupled cavity systems in which one of the cavities was interacting with a single qubit that was utilized to elevate the atom-field nonlinearity through the gain mechanism \cite{zheng2017nonreciprocal}. Around the same time, Miri et al. reported a unified framework to establish optical isolation and non-reciprocity in multimode optomechanical cavities \cite{miri2017optical}. Since then several studies have been conducted to analyze the breaking of time-reversal symmetry of light propagation in cavity quantum optomechanics (see, for example, Refs.~\cite{xu2020quantum, liu2019nonreciprocal,  lan2022nonreciprocal}).

One particularly important study in this context was carried out by L\"u et al. where they focused on a pump-probe driven fiber coupled optomechanical ring resonator which was capable of spinning \cite{lu2017optomechanically}. Under steady-state conditions, L\"u et al. were able to theoretically show that with the aid of rotational Sagnac effect not only is non-reciprocal probe light propagation possible to achieve but additionally the dispersion properties in the optomechanically induced transparency (OMIT) region allows one to achieve slow light propagation. More recently, we and others have further studied spinning ring resonator architectures and predicted irreversible refraction \cite{maayani2018flying}, better control of non-reciprocity with slow \& fast light propagation \cite{mirza2019optical}, nonreciprocal entanglement \cite{jiao2020nonreciprocal}, breaking Anti-$\mathscr{P}\mathscr{T}$-symmetry \cite{zhang2020breaking}, photon blockade \cite{jing2021nonreciprocal}, and phonon blockade \cite{yao2022nonreciprocal} via the control of the rotational Sagnac effect.

Motivated by the aforementioned studies, in this paper we have focused on the problem of qubit coupled spinning optomechanical ring resonators. Here we asked the question in what ways can the presence of a weakly coupled two-level QE impact the probe light transmission through such a hybrid architecture? In addition to being fundamentally an interesting problem, this study may find applications in non-reciprocal quantum circuitry \cite{bernier2017nonreciprocal} and quantum networking/internet \cite{kimble2008quantum} where qubits play a central role in the process of information storage and manipulation. As some of the key findings of this work, we notice (under a set of experimentally feasible parameters) that the presence of a weakly coupled qubit assists in realizing 3\% enhancement in the OMIT peak height, and the transparency window broadens by $1MHz$. These differences appear due to the opening of a back reflection channel for the probe light propagation which it is not able to attain without the presence of a QE. Additionally, we note that qubit results in a reduction in the group delay of the probe field compared to the no-qubit case. These results indicated that even a weakly coupled qubit can impact the probe light transmission considerably in spinning optomechanical resonators. 

The rest of the paper is organized as follows. In Section II we present the theoretical model and calculate the probe light transmission under steady-state conditions. In Section III, we present results focusing on non-reciprocal probe light propagation and controlling the group velocity of light (slow light propagation). Finally, in Section IV, we close with a summary of the main results and discuss possible future directions of this work.


\section{Theoretical description}
\subsection{Setup and Hamiltonian}
As shown in Fig.~1, we consider a hybrid setup of a spinning optomechanical resonator coupled with a stationary two-level QE (hereinafter also referred to as an atom or a qubit) residing at the center of the resonator. Under the rotating wave approximation, and in a frame rotating at a frequency $\omega_l$, the Hamiltonian of the system (with setting $\hbar=1$) can be decomposed into three parts
\begin{equation}
\label{eq:HTot}
{\bf \hat{\mathscr{H}}}=\hat{\mathcal{H}}_{0}+\hat{\mathcal{H}}_{int}+\hat{\mathcal{H}}_{dr},
\end{equation}
where $\hat{\mathcal{H}}_{0}$ represents the free/non-interacting part of the Hamiltonian which is given by
\begin{equation}
\hat{\mathcal{H}}_{0}=\sum_{\nu=a,b}\Delta_{c_\nu}\hat{\nu}^{\dagger}\hat{\nu}+\Delta_{eg}\hat{\sigma}^{\dagger}\hat{\sigma}+\frac{\hat{p}^{2}}{2m}+\frac{\hat{p}^{2}_{\theta}}{2m r^{2}}+\frac{1}{2}m\omega^{2}_{m}\hat{x}^{2}.
\end{equation}
The first term on the right hand side of $\hat{\mathcal{H}}_{0}$ describes the Hamiltonian of two counter-propagating optical modes $a$ and $b$ with annihilation operators $\hat{a}$ and $\hat{b}$, and frequencies $\omega_{c_a}$ and $\omega_{c_b}$, respectively. $\Delta_{c_\nu}:=\omega_{c_\nu}-\omega_{l}$ with $\omega_l$ being the frequency of the strong pump field.  Here we have assumed that the pump and probe fields only influence $\hat{a}$-mode and there is no backscattering between the optical modes. The second term represents a non-interacting atom with transition frequency $\omega_{eg}$ ($\Delta_{eg}:=\omega_{eg}-\omega_{l}$) and $\hat{\sigma}(\hat{\sigma}^{\dagger})$ the atomic lowering (raising) operator. $\hat{p}$ and $\hat{p}_{\theta}$ appearing in the third and fourth term show linear and angular momentum operators while $m$ is the mass and $r$ is the radius of the ring resonator. The last term characterizes the mechanical harmonic motion of the resonator in a breathing mode fashion due to strong pumping. The frequency of mechanical oscillation is $\omega_m$ and $\hat{x}$ is the position operator associated with this mechanical motion.

Proceeding further, we discuss the interaction part of the Hamiltonian which in our model takes the form 
\begin{equation}
\hat{\mathcal{H}}_{int}=\sum_{\nu=a,b}\Big(-g_{\nu}\hat{x}\hat{\nu}^{\dagger}\hat{\nu}+J_{\nu}(\hat{\nu}^{\dagger}\hat{\sigma}+\hat{\sigma}^{\dagger}\hat{\nu})\Big).
\end{equation}
$\hat{\mathcal{H}}_{int}$ is composed of two parts: the first part with a prefactor $g_\nu$ (coupling strength parameter) shows the typical nonlinear optomechanical coupling \cite{bowen2015quantum} between each optical mode and the mechanical degree of freedom. The second part (with coupling parameter $J_\nu$) is the standard Jaynes-Cummings interaction (with energy conserving terms) between the optical mode and the qubit extended to both modes (see Ref.~\cite{sundaresan2015beyond} for the experimental realization of the multimode Jaynes-Cummings model with energy conserving and non-conserving terms).
 
Next, in the following, we present the drive Hamiltonian as
\begin{equation}
\hat{\mathcal{H}}_{dr}=i\sqrt{\kappa^{(ex)}_{a}}\Big(\varepsilon_{l}\hat{a}^{\dagger}+\varepsilon_{p}\hat{a}^{\dagger}e^{-i\eta t}-h.c.\Big),
\end{equation}
where we have assumed both pump and probe fields are launched from the left end of the fiber and coupled with the $\hat{a}$ mode of the resonator with a coupling strength $\kappa^{(ex)}_a$. The frequency of the probe field is $\omega_p$ and the detuning between both fields is $\eta:=\omega_{p}-\omega_{l}$. Probe and pump field amplitudes ($\varepsilon_p $ and $\varepsilon_l $) are related to pump and probe power ($P_{in}$ and $P_{l}$), respectively through $
\varepsilon_p =\sqrt{{P_{in}/\hbar\omega_{p}}}$ and $\varepsilon_{l}=\sqrt{{P_{l}/\hbar\omega_{l}}}$. Non-vanishing commutation and anticommutation relations among different operators appearing in the total Hamiltonian ${\bf \hat{\mathscr{H}}}$ are given by: $[\hat{x},\hat{p}_{x}]=i,[\hat{\theta},\hat{p}_{\theta}]=i, [\hat{a},\hat{a}^{\dagger}]=1, [\hat{b},\hat{b}^{\dagger}]=1, \lbrace\hat{\sigma}^{\dagger},\hat{\sigma}\rbrace=1. $ \\
Finally, we point out that owing to the presence of angular momentum in our model, the resonator performs rotary motion (also referred to as rotation/spin here onwards). As a result of this rotation, the optical mode frequency undergoes {\it Sagnac-Fizeau} effect \cite{malykin2000sagnac} which shifts the bare frequency of both optical modes to
\begin{equation}\label{EqSagnac}
\begin{split}
&\omega_{c_{\nu}}\longrightarrow\omega_{c_{\nu}}+\Delta^{(\nu)}_{sag},\\
&\text{where}\hspace{2mm}\Delta^{(\nu)}_{sag}=\frac{n r\Omega\omega_{c_{\nu}}}{c}\Bigg(1-\frac{1}{n}-\frac{\lambda}{n}\frac{dn}{d\lambda}\Bigg).
\end{split}
\end{equation}
$\Delta^{\nu}_{sag}$, $n$ are the $\nu^{th}$ mode Sagnac-Fizeau shift and refractive index of the resonator, respectively. $c$ is the group velocity of light and $\Omega=d\theta/dt$ is the speed of rotation which is taken to be positive (negative) in the clockwise (counterclockwise) rotary direction. The $dn/d\lambda$ term represents the relativistic dispersion correction to the frequency shift which we are going to ignore in this work. Note that the first term in $\Delta^{\nu}_{sag}$ originates from the rotation of the resonator itself which is the Sagnac term while the last two terms display the Fizeau drag due to the light propagation through the medium of the moving resonator.

\subsection{Heisenberg-Langevin equations of motion}
The system dynamics can be described in the Heisenberg picture where the time evolution of the average values of the system operators can be readily calculated. For the model under study, we find
\begin{align}\label{eq:LangEqs}
&\frac{d\langle\hat{a}(t)\rangle}{dt}=-i(\Delta_{c_{a}}-i\beta_{a})\langle\hat{a}\rangle-g_{a}\langle\hat{x}\hat{a}\rangle-iJ_{a}\langle\hat{\sigma}\rangle\nonumber\\
&\hspace{15mm}+\sqrt{\kappa^{(ex)}_{a}}(\varepsilon_{l}+\varepsilon_{p}e^{-i\eta t}),\nonumber \\
&\frac{d\langle\hat{b}(t)\rangle}{dt}=-i(\Delta_{c_{b}}-i\beta_{b})\langle\hat{b}\rangle-g_{b}\langle\hat{x}\hat{b}\rangle-iJ_{b}\langle\hat{\sigma}\rangle,\nonumber\\
&\frac{d\langle\hat{\sigma}(t)\rangle}{dt}=-i\widetilde{\Delta}_{eg}\langle\hat{\sigma}\rangle+iJ_{a}\langle\hat{\sigma}_{z}\hat{a}\rangle+iJ_{b}\langle\hat{\sigma}_{z}\hat{b}\rangle,\nonumber\\
&\frac{d\langle\hat{\sigma_z}(t)\rangle}{dt}= 2iJ_{a}\left(\langle\hat{\sigma}^\dagger\hat{a}\rangle-\langle\hat{a}^\dagger\hat{\sigma}\rangle\right)+2iJ_{b}\left(\langle\hat{\sigma}^\dagger\hat{b}\rangle-\langle\hat{b}^\dagger\hat{\sigma}\rangle\right),\nonumber\\
&\frac{d^{2}\langle\hat{x}(t)\rangle}{dt^{2}}=-\Big(\omega^{2}_{m}+\gamma_{m}\frac{d}{dt}\Big)\langle\hat{x}\rangle+\frac{g_{a}}{m}\langle\hat{a}^{\dagger}\hat{a}\rangle+\frac{g_{b}}{m}\langle\hat{b}^{\dagger}\hat{b}\rangle\nonumber\\
&\hspace{16mm}+\frac{\langle\hat{p}^{2}_{\theta}\rangle}{m^{2}r^{3}},\nonumber\\
&\frac{d\hat{\langle\theta\rangle}(t)}{dt}=\frac{\langle\hat{p}_{\theta}\rangle}{mr^{2}},\hspace{7mm}\text{and}\hspace{7mm} \frac{d\langle\hat{p}_{\theta}(t)\rangle}{dt}=0.
\end{align}
Here $\widetilde{\Delta}_{eg}=\Delta_{eg}-i\gamma^{\ast}$. Note that in the above equations, we have phenomenologically added the atomic, optical, and mechanical dissipation terms where $\gamma^{\ast}$ is the spontaneous emission rate, $2\beta_{a}=\kappa^{(ex)}_{a}+\kappa^{(in)}_{a}$ and $2\beta_{b}=\kappa^{(ex)}_{b}+\kappa^{(in)}_{b}$ are the optical mode net leakage rates (for mode $\hat{a}$ and $\hat{b}$, respectively) while mechanical mode decay rate is given by $\gamma_{m}$. Note that while considering the mechanical motion we have neglected the thermal Langevin force (Hermitian Brownian noise) \cite{genes2008emergence} as it averages to zero. 

\subsection{Steady-state solutions}
The solution of the above Heisenberg-Langevin equations requires the equations of motion for atom-field correlated operators, for instance $\langle\hat{\sigma}_{z}\hat{a}\rangle$ and $\langle\hat{\sigma}^\dagger\hat{b}\rangle$. However, the set of Eq.~(\ref{eq:LangEqs}) is nonlinear, and in general, it is difficult to obtain an analytic solution. To overcome this issue we apply the mean-field approximation \cite{agarwal2010electromagnetically}. According to mean-field approximation we write $\langle \hat{x}\hat{a}\rangle \cong \langle \hat{x}\rangle \langle \hat{a}\rangle$, $\langle \hat{a}^{\dagger}\hat{a}\rangle \cong \langle \hat{a}^{\dagger}\rangle \langle \hat{a}\rangle$, and $\langle \hat{b}^{\dagger}\hat{b}\rangle \cong \langle \hat{b}^{\dagger}\rangle \langle \hat{b}\rangle$. Inserting these forms of the average values, we follow the standard procedure \cite{boyd2003nonlinear, jing2015optomechanically} and expand the expectation value of the arbitrary operator $\langle\hat{\mathcal{O}}(t)\rangle$ in its steady-state value and small fluctuating values around it as
\begin{equation}
\label{eq:SSansatz}
\langle\hat{\mathcal{O}}(t)\rangle = \mathcal{\widetilde{O}}+\delta \mathcal{O}_{-}e^{-i\eta t}+\delta \mathcal{O}_{+}e^{+i\eta t}.
\end{equation}
For the present problem we have $\hat{\mathcal{O}}(t) \in \lbrace \hat{a}(t), \hat{b}(t), \hat{\sigma}(t),\hat{x}(t) \rbrace$. Note that the operator $\hat{\sigma}_z$ will be taken care of through the weak-coupling regime of qubit-field interaction as discussed later in Sec.~III(B). Inserting Eq.~(\ref{eq:SSansatz}) into Eq.~(\ref{eq:LangEqs}) and assuming fluctuations to be much smaller than the mean values (i.e. $|\delta a_{\pm}|<<|\widetilde{a}|, |\delta b_{\pm}|<<|\widetilde{b}|, |\delta \sigma_{\pm}|<<|\widetilde{\sigma}|$, and $\lbrace \delta x, \delta x^{\ast}\rbrace<<|\widetilde{x}|$), we obtain the following expressions for the steady-state values of the average of the operators
\begin{equation}\label{EqsAvg}
\begin{split}
& \widetilde{a}=\bigg(\frac{-iJ_a}{i\Delta_{c_{a}}-ig_{a}\widetilde{x}+\beta_{a}}\bigg)\widetilde{\sigma}+\Bigg(\frac{\sqrt{\kappa^{(ex)}_{a}}\varepsilon_{l}}{i\Delta_{c_{a}}-ig_{a}\widetilde{x}+\beta_{a}}\Bigg),\\
&\widetilde{b}=\bigg(\frac{-iJ_{b}}{i\Delta_{c_{b}}-ig_{b}\widetilde{x}+\beta_{b}}\bigg)\widetilde{\sigma},\\
& \widetilde{\sigma}=\bigg(\frac{-J_{a}}{\widetilde{\Delta}_{eg}}\bigg)\widetilde{a}+\bigg(\frac{-J_{b}}{\widetilde{\Delta}_{eg}}\bigg)\widetilde{b},\\
& \widetilde{x}=\bigg(\frac{g_{a}}{m\omega^{2}_{m}}\bigg)\vert \widetilde{a}\vert^{2}+\bigg(\frac{g_{b}}{m\omega^{2}_{m}}\bigg)\vert \widetilde{b}\vert^{2}+r\bigg(\frac{\Omega}{\omega_{m}} \bigg)^{2}.
\end{split}
\end{equation}
Here $\vert\Omega\vert=\frac{d\theta}{dt}$ refers to the magnitude of the spinning rate. Similarly, we find the fluctuating part of the operators taking the form
\begin{equation}\label{EqsFluc}
\begin{split}
&(\beta_{a}+i\Delta_{c_{a}}-ig_{a}\widetilde{x}-i\eta)\delta a_{-}-ig_{a}\widetilde{a}\delta x=-iJ_{a}\delta \sigma_{-}\\
&\hspace{57mm}+\sqrt{\kappa^{(ex)}_{a}}\varepsilon_{p},\\
&(\beta_{a}-i\Delta_{c_{a}}+ig_{a}\widetilde{x}-i\eta)\delta a^{\ast}_{+}+ig_{a}\widetilde{a}^{\ast}\delta x=iJ_{a}\delta \sigma^{\ast}_{+},\\
&(\beta_{b}+i\Delta_{c_{b}}-ig_{b}\widetilde{x}-i\eta)\delta b_{-}-ig_{b}\widetilde{b}\delta x=-iJ_{b}\delta \sigma_{-},\\
&(\beta_{b}-i\Delta_{c_{b}}+ig_{b}\widetilde{x}-i\eta)\delta b^{\ast}_{+}+ig_{b}\widetilde{b}^{\ast}\delta x=iJ_{b}\delta \sigma^{\ast}_{+},\\
& (\eta+\widetilde{\Delta}_{eg})\delta \sigma_{-}=-J_{a}\delta a_{-}-J_{b}\delta b_{-},\\
& (\omega^{2}_{m}-\eta^{2}-i\eta\gamma_{m})\delta x=\frac{g_{a}}{m}\big(~\widetilde{a}^{\ast}\delta a_{-}+\widetilde{a}\delta a^{\ast}_{+}\big)\\
&\hspace{32mm}+\frac{g_{b}}{m}\big(~\widetilde{b}^{\ast}\delta b_{-}+\widetilde{b}\delta b^{\ast}_{+}\big).
\end{split}
\end{equation}
Solving the system of coupled linear equations given in Eq.~\eqref{EqsAvg} and Eq.~\eqref{EqsFluc}, in the next section we discuss the propagation properties of the probe field in the spectral domain.


\begin{table*}
\begin{tabular}{ p{8cm} p{3cm} p{2.35cm} }
\multicolumn{3}{c}{{\bf Table I:\label{tab:Tab1} Parameters and their values used in the results}} \\
\hline
\hline
{\bf Definitions} & {\bf Symbols} & {\bf Values} \\
 \hline\hline
Mass of the mechanical oscillator &$m$ & $2ng$\\
Mechanical oscillator frequency & $\omega_{m}$  & $200MHz$  \\
Mechanical damping rate & $\gamma_{m}$ & $0.2MHz$\\
Optomechanical coupling rate & $J_{a/b}=\omega_{c_{a/b}}/r$\\
\hline
Optical wavelength & $\lambda$ & $ 1.55\mu m$\\
Refractive index & $n$ & $1.44$\\
Optical resonant mode frequency & $\omega_{c_a/c_b}$  & $193.5 THz$\\
Quality factor of the optical resonator & $Q$ & $3\times 10^{7}$\\
Pump power & $P_{l}$ & $10W$\\
Resonator radius & $r$ & $0.25mm$\\
Speed of light in the optical medium & $v =3\times 10^{8}/n$\\
Cavity waveguide coupling rate & $\kappa^{(ex)}_{a/b} =\omega_{c}/Q$\\
\hline
Atom-cavity coupling rate & $ J_{a/b}$ & $0.75\kappa^{(ex)}$\\
Atom-cavity detuning & $ \Delta_{eg}$ & $0.5\omega_{m}$\\
Spontaneous emission rate & $\gamma^{\ast}$ & $0.03 MHz$\\
\hline
\end{tabular}
\end{table*}

\subsection{Probe field transmission and reflection rates}
In the forward and backward directions, we respectively define two output operators $\hat{a}_{out}$ and $\hat{b}_{out}$ to describe the probe field transmission and reflection rates. These output operators are linked with the intracavity optical field and input operators through the standard Collett and Gardiner input-output relations \cite{gardiner1985input}
\begin{subequations}
\begin{eqnarray}
\label{eq:InOut}
\hat{a}_{out}=\hat{a}_{in}-\sqrt{\kappa^{(ex)}}\delta a_{-},\\
\hat{b}_{out}=\hat{b}_{in}-\sqrt{\kappa^{(ex)}}\delta b_{-},
\end{eqnarray}
\end{subequations}
where for simplicity we have selected a symmetric coupling between the fiber and the ring resonator i.e.  $\kappa^{(ex)}_{a}=\kappa^{(ex)}_{b}=\kappa^{(ex)}$. For a coherent state-driven system we can replace $\hat{a}_{in}$ by the average value of the input probe field i.e. $\langle \hat{a}_{in}\rangle=\varepsilon_{p}$. Hence we define the transmission $T$ and reflection rates $R$ for the probe field in the following manner
\begin{subequations}
\begin{eqnarray}
\label{eq:T&R}
T:=\frac{\langle \hat{a}^{\dagger}_{out}\hat{a}_{out}\rangle} {\langle \hat{a}^{\dagger}_{in}\hat{a}_{in}\rangle}=\Bigg|1-\frac{\sqrt{\kappa^{(ex)}}}{\varepsilon_{p}}\delta a_{-}\Bigg|^{2},\\
R:=\frac{\langle \hat{b}^{\dagger}_{out}\hat{b}_{out}\rangle} {\langle \hat{a}^{\dagger}_{in}\hat{a}_{in}\rangle}=\Bigg|-\frac{\sqrt{\kappa^{(ex)}}}{\varepsilon_{p}}\delta b_{-}\Bigg|^{2}.
\end{eqnarray}
\end{subequations}
Note that in the backward direction (reflection rate) the input drive is a vacuum state; therefore, we utilized $\langle \hat{a}^{\dagger}_{in}\hat{a}_{in}\rangle$ as the normalization factor.

\section{Results}
\subsection{Parameters}
In Table I we summarize the parameters used to plot the results. For the spinning optomechanical part of the problem we followed the parameters mentioned in Ref. \cite{lu2017optomechanically, maayani2018flying, mao2022experimental}, while for the cavity QED part our parameters are quite close to what is reported in Ref. \cite{akram2015tunable}. For simplicity we have assumed both optical modes to be identical in their frequencies as well as in their interaction with mechanical, atomic and environmental part of the setup (a fully symmetric case) i.e. we set  $g_{a}=g_{b}=g$, $J_{a}=J_{b}=J$, $\kappa^{(in)}_{a}=\kappa^{(in)}_{b}=\kappa^{(in)}$. 
Furthermore, we consider an on-resonance condition for cavity QED meaning $\omega_{eg}=\omega_c\implies\Delta_{eg}=\Delta_{c}$ and assume internal and external loss of the optical mode to be the same i.e. $\kappa^{(in)} =\kappa^{(ex)}$.  Finally, to distinguish between weak and strong-coupling between the emitter and optical modes we introduce the cooperativity factor $\mathcal{C}:=J^{2}/(2\kappa_{ex}\gamma^{\ast})$. This single dimensionless parameter $\mathcal{C}$ distinguishes between the weak coupling regime and strong coupling regime of cavity QED for which $\mathcal{C}<1$ and $\mathcal{C}>1$, respectively \cite{cui2005quantum, auffeves2010controlling}. In the following (and for the rest of this work) we focused on the weak coupling regime and selected $\mathcal{C}=0.5$. We note that this value of cooperativity is experimentally feasible and lies well below the cooperativity values achieved in some of the recent cavity QED experiments (see for instance Ref.~\cite{samutpraphoot2020strong} where $\mathcal{C}\cong71$ has been reported in the cavity QED platform based on $^{87}Rb$ atoms coupled with a photonic crystal cavity.
 
\subsection{Weak coupling regime of cavity QED}
Next, to find a closed form solution of Eq. set (\ref{eq:LangEqs}) we apply the commonly adopted ``weak-excitation assumption" \cite{waks2006dipole, fan2010input} for the qubit. Under this assumption, we assume the qubit is weakly excited such that throughout the dynamics one can set $\langle\hat{\sigma}_{z}\rangle\cong -1$. Physically, the validity of this assumption relies on the weak coupling of the qubit with the cavity mode field, which we have assumed here (see Table I and compare values of $J_{a/b}$ in units of $\kappa^{(ex)}$ and $\gamma^\ast$). The weak-excitation assumption allows us to ignore dynamics of $\langle\hat{\sigma}_z(t)\rangle$ and simplify the equation of motion for $\langle\hat{\sigma}(t)\rangle$ to
\begin{equation}
\frac{d\langle\hat{\sigma}(t)\rangle}{dt}\cong-i\Delta_{eg}\langle\hat{\sigma}\rangle-ig_{a}\langle\hat{a}\rangle-ig_{b}\langle\hat{b}\rangle,
\end{equation}
which in turn helps us to arrive at the equation sets given in Eq.~\eqref{EqsAvg} and Eq.~\eqref{EqsFluc} which are then simultaneously solved to obtain $T$ and $R$ rates.

\subsection{Non-reciprocal light propagation}
\begin{figure}
\centering
\begin{tabular}{@{}cccc@{}}
\includegraphics[width=3.05in, height=2.15in]{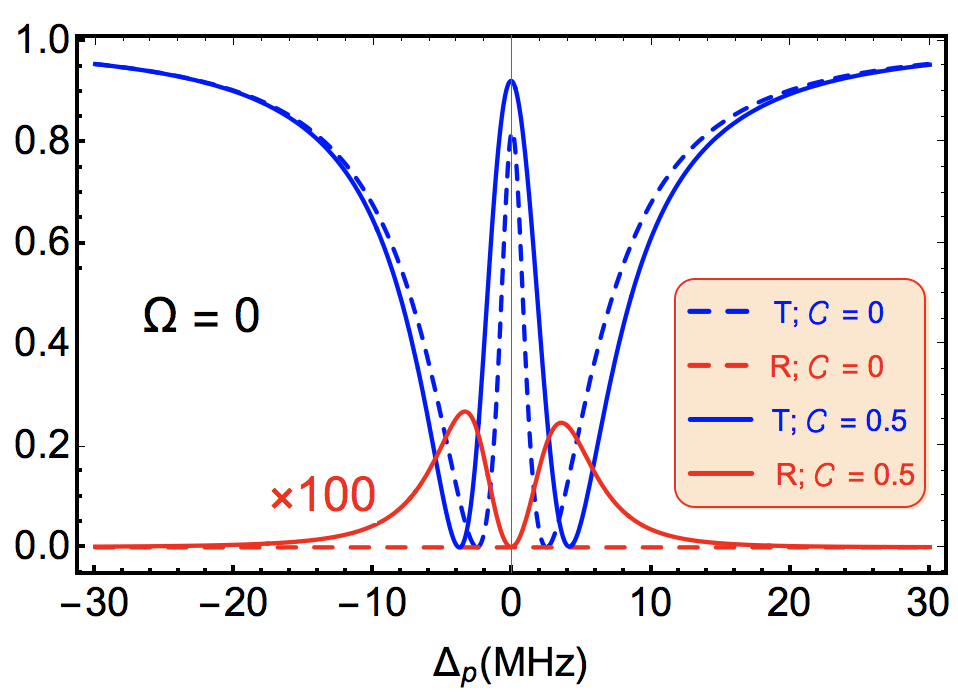} 
\end{tabular}
\captionsetup{
format=plain,
margin=1em,
justification=raggedright,
singlelinecheck=false
}
\caption{(Color online) Probe light transmission $T$ and reflection rate $R$ for a single non-spinning optomechanical ring resonator coupled with a resonant two-level QE (i.e. $\omega_{eg}=\omega_{c_{a}}=\omega_{c_{b}}$). Parameters for this and all other plots are taken from Table~1. Both blue and red dashed (solid) curves correspond to the situations in which QE is decoupled from (coupled with) the optical mode. In the coupled cases, the cooperativity value $\mathcal{C}=0.5$ is set to be in the weak coupling regime such that the assumption of the low excitation limit is justified. Note that the red solid curve has been magnified by a factor of $100$ to fit the scale of the plot.}\label{Fig2}
\end{figure}

To set the stage for our results, we begin from Fig.~\ref{Fig2} where we ignore the spin degree of freedom completely and plot the probe light transmission and reflection rates for the cases of no coupling ($\mathcal{C}=0$, dashed curves) and weak coupling ($\mathcal{C}=0.5$, solid curves) between the QE and optical field. In the absence of qubit-field interaction we obtain
\begin{align}
\delta a_-=-\frac{\sqrt{\kappa^{(ex)}} \varepsilon_p\lbrace ig^2_a|a|^2+m\widetilde{\beta}_-\Gamma_m \rbrace}{ig^2_a|a|^2\widetilde{\beta}_-\widetilde{\beta}^\ast_-\lbrace ig^2_a|a|^2+m\widetilde{\beta}_-\Gamma_m \rbrace},
\end{align}
where $\widetilde{\beta}_-+i\eta=(\beta_a-i\Delta_c+ix g_a)$ and $\Gamma_m=\omega_m-i\eta(\gamma_m-\eta)$. As evident from the red dashed curve of Fig.~\ref{Fig2}, in this case, the reflection rate $R$ vanishes for all values of $\Delta_p$ due to the absence of any mechanism that can route the photons in the clockwise/backward direction in the ring resonator (see Fig.~\ref{Fig1}). The $T$ curve on the other hand exhibit the standard optomechanically induced transparency (OMIT) behavior \cite{weis2010optomechanically} with the peak residing at the resonance point $\Delta_p=0$ and the peak width given by $\gamma_m+g^2_a|a|^2/(m^2\omega^2_m\beta )\sim 2MHz$. On the sides of the OMIT peak, we notice two minima at $\Delta_p=\pm 1.5MHz$ indicating the complete absorption of probe light at these off-resonant frequencies. 

\begin{figure*}
\centering
\begin{tabular}{@{}cccc@{}}
\includegraphics[width=3.15in, height=2.15in]{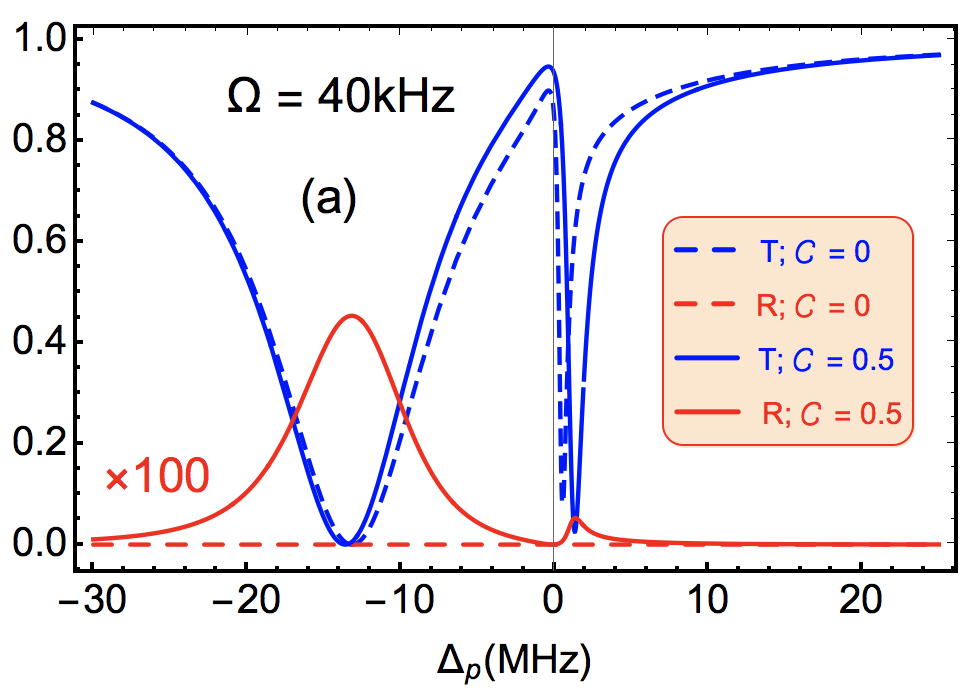} &
\hspace{4mm}\includegraphics[width=3.15in, height=2.15in]{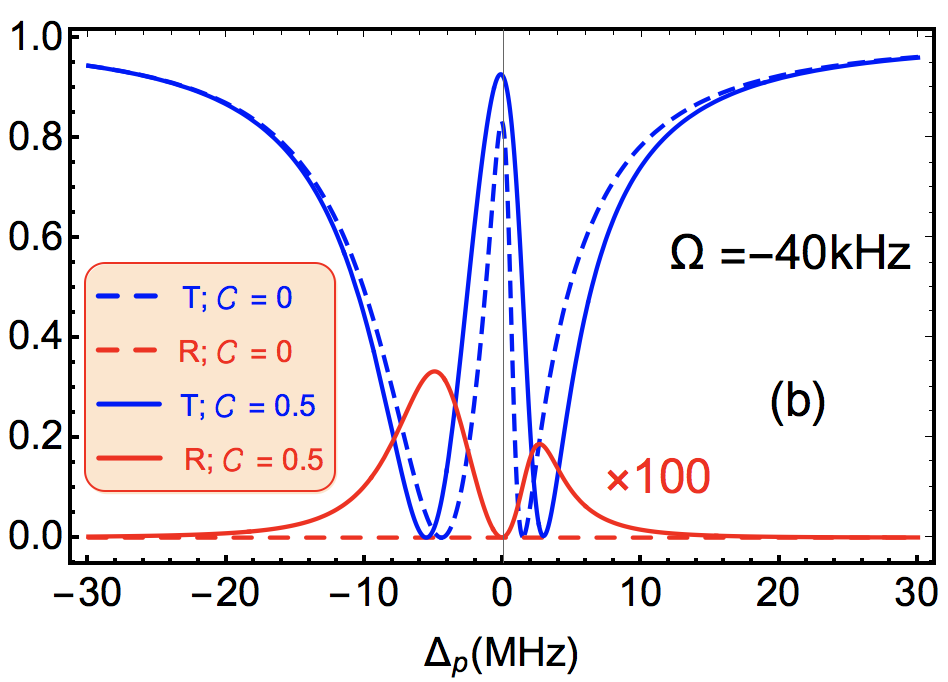} 
\end{tabular}
\captionsetup{
format=plain,
margin=1em,
justification=raggedright,
singlelinecheck=false
}
\caption{(Color online) Probe light transmission and reflection rates for a (a) clockwise, and a (b) counterclockwise spinning resonator. In both cases the magnitude of spinning rate $|\Omega|$ is set equal to $40kHz$. Again we have magnified the solid red reflection curves to match the scale of the plot. }\label{Fig3}
\end{figure*}

Next, in Fig.~\ref{Fig2} we introduce non-zero cooperativity between the QE and the optical field. Our results indicate two key behaviors. One being the emergence of a small yet non-zero reflection ($R_{max}\sim 0.2\%$) rate around $\Delta_p=0$ point. We emphasize that unlike the setup considered by L\"u et al. \cite{lu2017optomechanically} (which guaranteed a unidirectional light propagation), the presence of a qubit breaks the one-way light propagation and opens a new reflection channel. Secondly, we notice that the qubit increases the OMIT peak by $\sim 3\%$ and aids to broaden the transparency window from $2MHz$ to $3MHz$. It is worthwhile to point out that it is known that the backscattering loss induced by material defects in the optical resonators always suppresses the OMIT effect (see for example \cite{svela2020coherent}). However, our results, on the other hand, indicate that qubit assists in the enhancement of OMIT and opens the possibility of OMIT engineering. Later in this section, we discuss how these effects can impact the dispersion properties of the transmitted probe light.

Proceeding further, we now include the spin degree of freedom in our study and plot $T$ and $R$ rates with $\Omega=40kHz$ and $\Omega=-40kHz$ in Fig.~\ref{Fig3}(a) and Fig.~\ref{Fig3}(b), respectively. In the absence of qubit coupling, the values of $\langle \hat{x}\rangle$ and $\langle \hat{a}\rangle$ are known to obey the following coupled steady-state values \cite{mirza2019optical}
\begin{align}\label{xass}
x=\frac{g^2_a|a|^2+mr\Omega^2}{m\omega^2_m},~~\text{and}~~ a=\frac{\sqrt{\kappa^{(ex)}}\varepsilon_p}{\beta+i\Delta_c-ig_a x}.
\end{align}
\begin{figure}
\centering
\begin{tabular}{@{}cccc@{}}
\hspace{-1mm}\includegraphics[width=3.15in, height=2.05in]{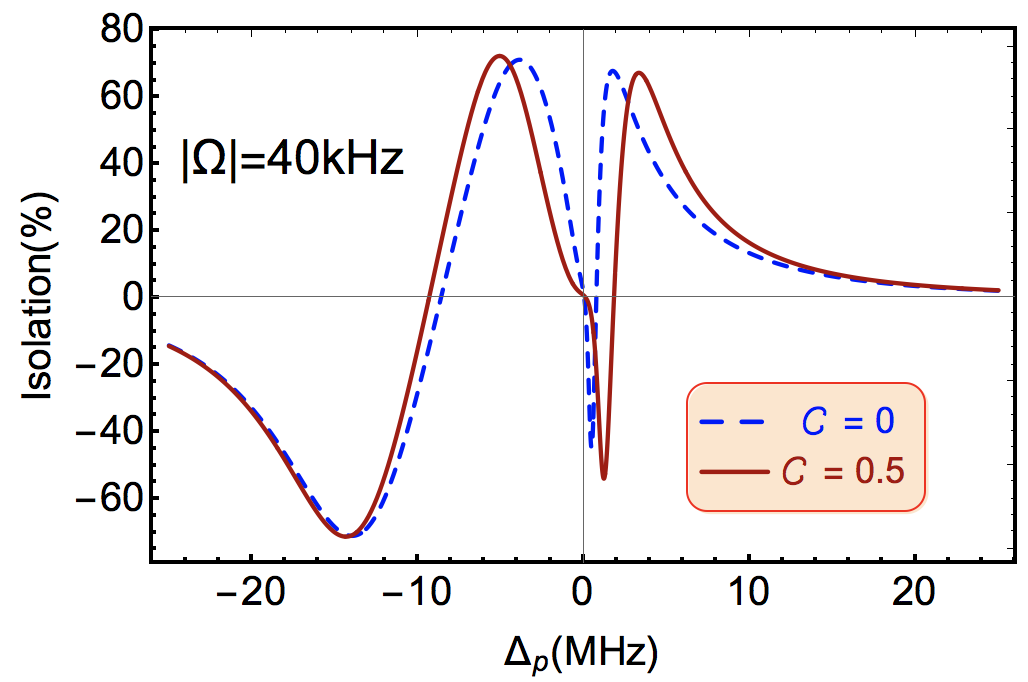} 
\end{tabular}
\captionsetup{
format=plain,
margin=1em,
justification=raggedright,
singlelinecheck=false
}
\caption{(Color online) Isolation $\mathcal{I}$ as a function of probe light detuning $\Delta_p$ for $\mathcal{C}=0$ and $\mathcal{C}\neq 0$ cases. The magnitude of the spinning rate for both cases has been fixed to $|\Omega|=40kHz$.}\label{Fig4}
\end{figure}

From Eqs.~\eqref{xass} and Eq.~\eqref{EqSagnac}, we notice that a non-zero spinning rate changes both $x$ and frequency shift due to the Sagnac effect which in turn can introduce the possibility of non-reciprocal light propagation without the need for the traditionally adopted magneto-optical effects. As a result, under the $\mathcal{C}=0$ case (blue dashed curves in Fig.~\ref{Fig3}), we observe that the $T$ near resonance alters considerably. 
\begin{figure*}
\centering
\begin{tabular}{@{}cccc@{}}
\includegraphics[width=3.1in, height=2.1in]{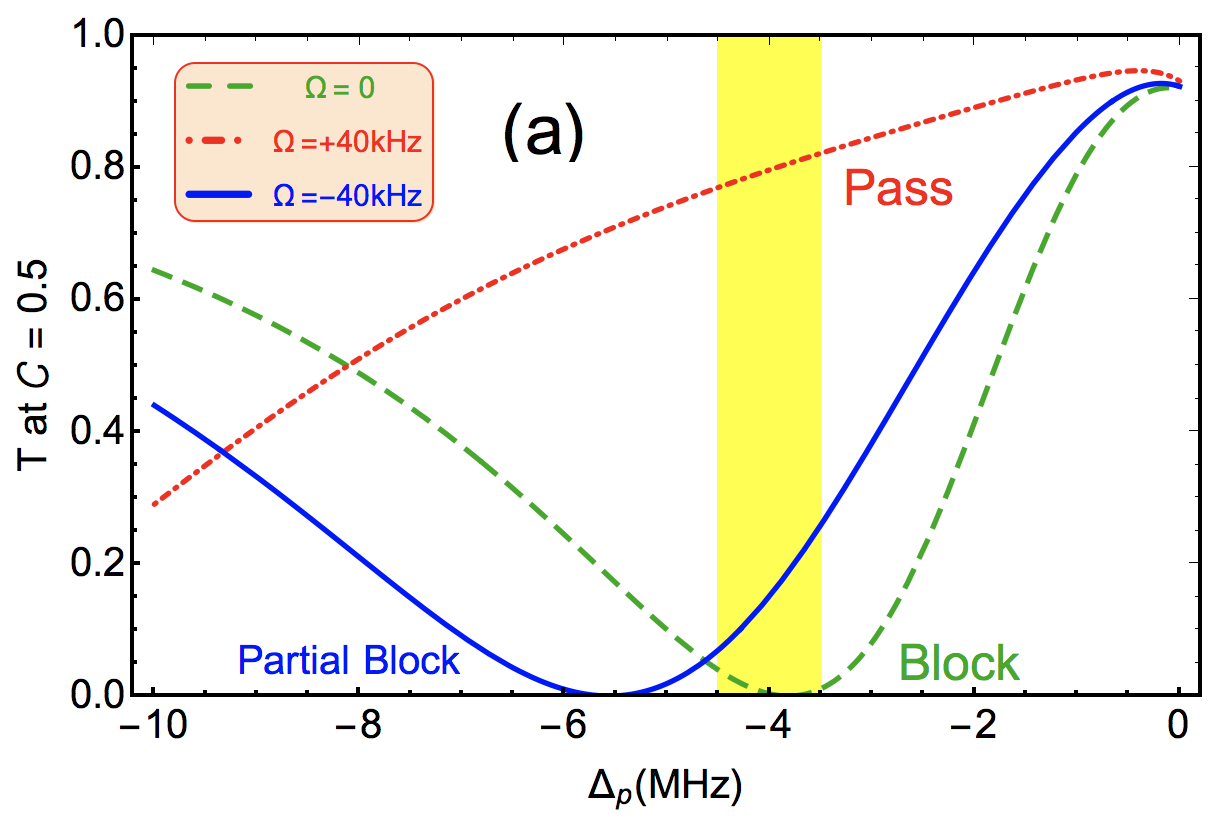} &
\hspace{3mm}\includegraphics[width=3.15in, height=2.05in]{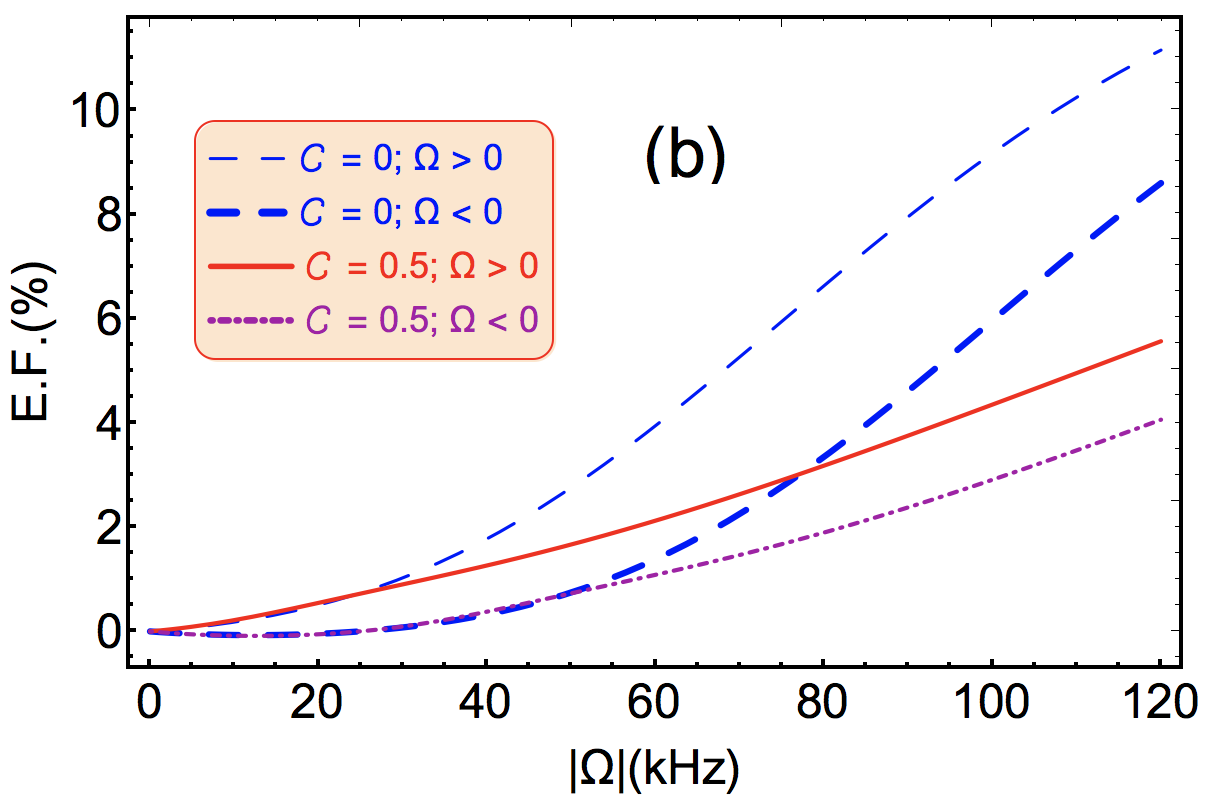} 
\end{tabular}
\captionsetup{
format=plain,
margin=1em,
justification=raggedright,
singlelinecheck=false
}
\caption{(Color online) (a) Probe light transmission rate as a function of probe detuning $\Delta_p$ in the selected range of $-10MHz\leq \Delta_p\leq 0$. The yellow highlighted region is to focus on the phenomenon of non-reciprocal light propagation. For all curves $\mathcal{C}=0.5$ is selected. (b) Transmission rate enhancement factor as a function of the magnitude of spinning rate $|\Omega|$ for $\mathcal{C}=0$ and $\mathcal{C}\neq 0$ cases.} \label{Fig5}
\end{figure*}

In particular, around $\Delta_p=-2MHz$ (where for $\Omega=0$ case we noticed an almost perfect absorption of probe light) now for $\Omega>0$ case $T\gtrsim 80\%$ while for $\Omega<0$ case the $T\gtrsim 50\%$. The presence of the qubit in this case ($\mathcal{C}\neq 0$) allows probe photons to be reflected with the maximum value of 0.4\% achieved for $\Omega>0$ around $\Delta_p=-5MHz$. Additionally, we notice that qubit assists in achieving a higher value of transmission at $\Delta_p=-2MHz$ for clockwise rotary direction, thereby providing additional means (besides the spin degree of freedom and $\Delta_p$) to control non-reciprocal transmission. Qualitatively, this slight enhancement in $T$ values at and around $\Delta_p\cong-2MHz$ point can be attributed to the possible destructive interference between the incoming and reflected probe light amplitudes. To see the impact of a qubit on the difference between clockwise transmission ($T_{cw}$) and counterclockwise transmission ($T_{ccw}$), in Fig.~\ref{Fig4} we plot the probe isolation $\mathcal{I}$ defined as \cite{zhang2020breaking}
\begin{align}
\mathcal{I}=|T_{cw}-T_{ccw}|,
\end{align}
where $T_{cw}$ and $T_{ccw}$ are normalized. We observe that the isolation for $\mathcal{C}\neq 0$ takes slightly higher and shifted maximum values as compared to the $\mathcal{C}=0$ case in the  $-10MHz\leq\Delta_p\leq 10MHz$ region enabling better non-reciprocity in the presence of qubit.


To further emphasize the non-reciprocal light propagation in the presence of qubit-optical field coupling ($\mathcal{C}\neq0$), in Fig.~\ref{Fig5}(a) we magnified the frequency region near $\Delta_p=-4MHz$ (yellow highlighted region) and plotted $T$ as a function of $\Delta_p$. We notice for the non-spinning case the transmission is completely blocked at this detuning. However, with the inclusion of qubit, the transmission at this frequency rises and can be further controlled by the spin direction. In particular, we noticed as we go from $\Omega<0$ to $\Omega>0$ case the $T$ rises from $\sim 20\%$ (partial block) to $\sim 80\%$ (considerable block), respectively. 

Next, to quantify the increase in $T$ on resonance for different spinning directions, we introduce the transmission rate enhancement factor E.F. which is defined at the probe detuning $\Delta_p = 0$ as
\begin{align}
E.F.=\frac{T\left(\Omega\neq 0)-T(\Omega = 0\right)}{T\left(\Omega = 0\right)}.
\end{align}
In Fig.~\ref{Fig5}(b) we plot the E.F. for both $\mathcal{C}=0$ and $\mathcal{C}\neq 0$ cases. In the absence of qubit, we compare the blue thick and thin dashed curves and notice that the transmission can be enhanced by $3\%$ by fixing the magnitude of spinning rate $\Omega$ to $120kHz$ but altering the spinning direction from counterclockwise to clockwise. This behavior extends down to the case when the qubit is present, however, the increase in the E.F. takes a value of $\sim 0.8\%$ (compare solid red and dotted-dashed magenta curves in Fig.~\ref{Fig5}(b)).

\begin{figure}
\centering
\begin{tabular}{@{}cccc@{}}
\hspace{-1mm}\includegraphics[width=3.15in, height=2.15in]{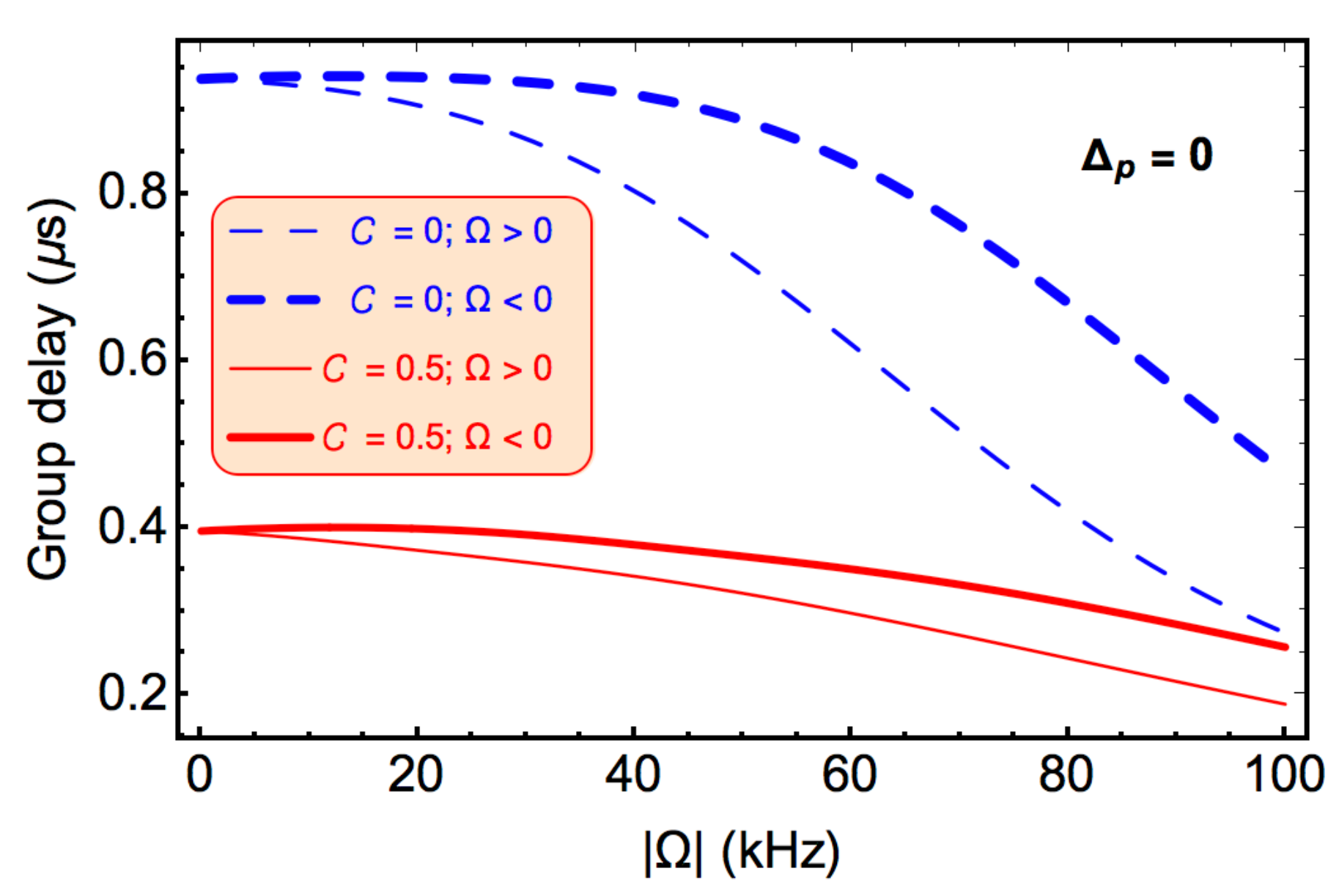}  
\end{tabular}
\captionsetup{
format=plain,
margin=1em,
justification=raggedright,
singlelinecheck=false
}
\caption{(Color online) Comparison of group delays (in $\mu s$ units) for the probe light transmission as a function of different spinning rates $|\Omega|$. For this plot, we have chosen the probe to be on resonance i.e. $\Delta_p=0.$}\label{Fig6}
\end{figure}
\subsection{Group delay and slow light propagation}
To address the possibility of slow or fast light propagation with possible applications in the optical storage of quantum information \cite{krauss2008we}, we define the group delay as 
\begin{align}
\tau_g=\frac{d{\rm arg}\left(t_p\right)}{d\Delta_p}.
\end{align}
In Fig.~\ref{Fig6} we plot $\tau_g$ as a function of the magnitude of spinning rate $|\Omega|$ for fixed probe detuning of $\Delta_p=0$. Similar to Fig.~\ref{Fig5}(b), in Fig.~\ref{Fig6} we have plotted curves for $\mathcal{C}=0$ and $\mathcal{C}\neq 0$ cases and focused on two different spinning directions for each case. We note that at $|\Omega|=0$, group delay takes more than twice the value for the $\mathcal{C}=0$ case compared to $\mathcal{C}\neq 0$ case. However, as soon as the $|\omega|>0$, group delay for all cases shows a decreasing trend. From Fig.~\ref{Fig6} and \ref{Fig3} we notice that at $|\Omega|=40kHz$ and $\Delta_p=0$ for clockwise spinning direction and in the absence of the qubit one can not only obtain more than $80\%$ transmission but a group delay of $0.8\mu$s can also be achieved. The qubit on the other hand, for the same values of $\Delta_p$ and $\Omega$, increases transmission slightly but allows slowing of light only up to $0.4\mu$s.

%

\section{Conclusions and Discussion}
In this work, we have studied the influence of a two-level quantum emitter on the probe light transmission and reflection properties (non-reciprocity and slow \& fast light) in hybrid spinning optomechanical ring resonators. Under the low excitation assumption and mean field approximation, we focused on the steady state behavior of probe light while QE remained weakly coupled with the optical mode of the cavity. In particular, we found that (under experimentally feasible parameters summarized in Table I) the presence of a qubit can open a back reflection channel that can aid OMIT peak enhancement by 3\% and a width increase up to 1MHz. Furthermore, the combination of a weakly coupled qubit and rotation Sagnac effect allows controlling the one-way light propagation by changing the qubit-field cooperativity parameter. For instance, at $\Delta_p=-4MHz$ no spinning, counterclockwise spinning at $|\Omega|=40kHz$, and clockwise spinning direction at $|\Omega|=40kHz$ can completely block, pass, and partially block the probe light transmission, respectively. Finally, we observed that qubit-field coupling (despite being assistive in increasing the transmission) degrades the group delay by an almost 1/2 factor as compared to the no-qubit case. All of these results indicate better control and considerable impact of even a weakly coupled QE on the non-reciprocal and dispersive probe light transmission in hybrid spinning microring resonators.

Due to the interplay of several degrees of freedom, the hybrid quantum system we analyzed in this paper becomes quite rich and allows us to explore disparate novel avenues. For instance, in our model, we have considered optical mode-mediated coupling between QE and mechanical vibrations. However, recent studies have also considered the possibility of direct coupling between the QE and mechanical motion \cite{o2010quantum, wu2016polarons}. Therefore, one possible extension of this work would be to examine a full hybrid model \cite{restrepo2017fully} in which QE, mechanical motion, and optical field all are directly and mutually coupled. Another extension would be to go beyond the weak interaction regime between the QE and optical field and to consider the possible applications of the strong atom-field generated effects (such as Rabi splitting and highly entangled atom-photon states \cite{larson2021jaynes}) in the field of quantum information science. One could also introduce a direct drive to the qubit which can provide an effective gain mechanism to the system and might lead to the formation of an optical nonreciprocal amplifier \cite{lin2019nonreciprocal}. We leave these and related problems as the possible future directions of this work.

\acknowledgments
We thank Dr. Wenchao Ge for the valuable discussions. Financial support for this work was provided by the National Science Foundation Grants No. 1757575 (Miami University Physics Department Research Experience for Undergraduates program), LEAPS-MPS 2212860, and Miami University College of Arts \& Science and Physics Department start-up funding.

\bibliographystyle{ieeetr}
\bibliography{paper}
\end{document}